\documentclass[11pt]{article}

\usepackage{epsf}
\usepackage{graphicx}

\newcommand{\lsim}{\raisebox{0.6mm}{$\, <$} \hspace{-3.0mm}\raisebox{-1.5mm}{\em $\sim \,$}}
\newcommand{\gsim}{\raisebox{0.6mm}{$\, >$} \hspace{-3.0mm}\raisebox{-1.5mm}{\em $\sim \,$}}

\newcommand{\beq}{\begin{equation}}   
\newcommand{\eeq}{\end{equation}}
\newcommand{\bea}{\begin{eqnarray}}   
\newcommand{\eea}{\end{eqnarray}}
\def\GEV#1{10^{#1}{\rm\,GeV}}

\def\gev{{\rm\,GeV}}

\def\mgluino{m_{\tilde{g}}}

\def\lrf#1#2{ \left(\frac{#1}{#2}\right)}
\def\lrfp#1#2#3{ \left(\frac{#1}{#2} \right)^{#3}}

\def\stau{\tilde{\tau}}
\def\cos{{\rm cos}}
\def\sin{{\rm sin}}
\def\tan{{\rm tan}}
\let\bar\overline 
\let\tilde\widetilde

\begin{document}
\baselineskip=18pt

\begin{titlepage}

\begin{flushright}
UT--10--15\\
IPMU--10--0137\\
\end{flushright}

\vskip 1.35cm
\begin{center}
{\Large \bf
Probing High Reheating Temperature Scenarios\\
at the LHC with Long-Lived Staus
}
\vskip 1.2cm
Motoi Endo$^{1,2}$, Koichi Hamaguchi$^{1,2}$, and Kouhei Nakaji$^{1}$
\vskip 0.4cm

{\it $^1$  Department of Physics, University of Tokyo,
Tokyo 113-0033, Japan\\
$^2$ Institute for the Physics and Mathematics of the Universe, 
University of Tokyo,\\ Chiba 277-8568, Japan
}

\vskip 1.5cm

\abstract{ 
We investigate the possibility of probing 
high reheating temperature scenarios at the LHC,
in supersymmetric models where the gravitino is the lightest supersymmetric particle, 
and the stau is the next-to-lightest supersymmetric particle.
In such scenarios, the big-bang nucleosynthesis and the gravitino abundance 
give a severe upper bound on the gluino mass.
We find that, if the reheating temperature is $\sim \GEV{8}$ or higher, 
the scenarios can be tested at the LHC 
with an integrated luminosity of ${\cal O}(1~\mathrm{fb}^{-1})$ at $\sqrt{s}=7$ TeV
in most of the parameter space.
}
\end{center}
\end{titlepage}

\setcounter{footnote}{0}
\setcounter{page}{2}

\section{Introduction}

Reheating temperature $T_R$ is one of the most important physical quantities 
in the cosmology, which represents the highest temperature of the universe
at the beginning of the radiation--dominated epoch.
From the viewpoint of cosmological observations, $T_R$ is hardly restricted;
the only constraint is  $T_R \gsim 5$~MeV in order to realize 
a successful big-bang nucleosynthesis (BBN)~\cite{TR5MeV}.\footnote{
The reheating temperature may be probed with gravitational wave background 
in future experiments~\cite{Nakayama:2008wy}. 
The BBN prediction on the $^6$Li abundance, 
affected by long-lived charged particles, may also be used as a probe
of the reheating temperature~\cite{Takayama:2007du}.}
On the other hand, from the viewpoint of the particle physics, $T_R$ has various 
implications. 
First of all, it is determined by the inflaton decay rate $\Gamma_\phi$.  
If interactions between the inflaton and lighter particles are
suppressed by the Planck scale $M_P=2.4\times \GEV{18}$ such as 
$\Gamma_\phi \sim m_\phi^3 / M_\mathrm{P}^2$, 
the reheating temperature becomes
$T_R \sim  \sqrt{\Gamma_\phi M_\mathrm{P}}
\sim \GEV{10} (m_\phi / \GEV{13} )^{3/2}$.
In some inflation models, typical values of 
the inflaton mass $m_\phi \sim \GEV{(9-15)}$ provide $T_R \sim \GEV{(4-13)}$ 
(cf.~\cite{Asaka:1999jb,Endo:2007sz}).
Secondly, the baryon asymmetry of the universe must be generated between the inflationary 
era and the BBN epoch, and
baryogenesis mechanisms give constraints on $T_R$.
For instance, the thermal leptogenesis 
scenario with heavy right--handed neutrinos~\cite{FY}
requires $T_R > {\cal O}(\GEV{9})$~\cite{Lepto_review}. 
Thirdly, in supersymmetric models, the gravitinos are produced in the thermal plasma. 
Thus, the so--called gravitino problems impose severe upper bounds on the reheating 
temperature~\cite{KKMY}.

In this letter, we discuss the possibility of probing high reheating temperature scenarios at the LHC.\footnote{%
Possibilities of constraining $T_R$ at colliders in gravitino/axino LSP scenarios have been discussed 
in Ref.~\cite{Choi:2007rh,Steffen:2008bt}, though detailed collider studies have not been performed.}
In particular, we consider the supersymmetric (SUSY) models with conserved R-parity,
where the gravitino is the lightest supersymmetric particle, 
and the stau is the next-to-lightest supersymmetric particle.\footnote{%
LHC signatures of such a scenario have been studied in relation to cosmology
from different points of view~\cite{Feng:2004mt}.}
We further assume that there is no entropy production after the inflation.

In such a scenario, the BBN and the gravitino abundance give an upper bound on 
the gluino mass~\cite{Fujii:2003nr,Pradler:2006qh,Ros}. As the reheating temperature is larger, 
the mass bound becomes severer, and thus the production cross section of the SUSY events 
is enhanced. The essential feature of the events is the existence of heavy charged tracks of 
long-lived staus. Since the staus tend to have a high transverse momentum and a low velocity, 
the events are almost background free, and they can be easily discovered at the LHC. 

In this framework, we will see that if the reheating temperature is $\sim \GEV{8}$ or higher, 
the scenarios can be tested even in the early stage of the LHC experiments, 
i.e. with an integrated luminosity of ${\cal O}(1~\mathrm{fb}^{-1})$ at $\sqrt{s}=7$ TeV
in a wide parameter space of the SUSY Standard Model (SSM). Furthermore, we will find 
that a large part of the parameter region which is consistent with the thermal leptogenesis 
is accessible at ${\cal O}(10~\mathrm{fb}^{-1})$ at $\sqrt{s}=14$ TeV.

\section{Cosmological Constraints}
In this section, we study the cosmological constraints on the present scenario. 
In Sec.~\ref{sec:Omega32}, it is shown that the gravitino abundance gives
upper bounds on the gaugino masses once a gravitino mass and a reheating 
temperature are provided.
In Sec.~\ref{sec:BBN}, we discuss the BBN constraints on the long-lived stau and
derive an upper bound on the gravitino mass for a given stau mass.
Combining these two constraints, we obtain upper bounds on the gluino mass, 
depending on the stau mass and the reheating temperature.

\subsection{Gravitino abundance}
\label{sec:Omega32}
The present energy density of the gravitino must be smaller than the 
observed energy density of the dark matter. 
The gravitino energy density from the thermal scattering is given
by~\cite{Moroi:1993mb,Bolz:2000fu,Pradler:2006hh}\footnote{
Eq.~(\ref{gluinomass}) potentially includes an ${\cal O}(1)$ 
uncertainty~\cite{Bolz:2000fu,Rychkov:2007uq}.
In the present scenario, the gravitinos are also produced from the late-time stau decay.
However, its contribution is negligible in the parameter region of our interest, and 
we neglect it in the following discussion.
The gravitinos are also directly produced by the inflaton decay~\cite{Endo:2007sz}, but 
we do not include its contribution since it depends on the inflation models.}
\begin{eqnarray}\label{Ythermal}
\Omega_{3/2}h^2 &\simeq&
\lrf{1\gev}{m_{3/2}}
\lrf{T_R}{\GEV{8}}
\nonumber\\
&& \times
\left[ 
0.14\lrfp{m_{\tilde{B}}}{1{\rm\,TeV}}{2}
+
0.38\lrfp{m_{\tilde{W}}}{1{\rm\,TeV}}{2}
+
0.34\lrfp{\mgluino}{1{\rm\,TeV}}{2}
\right]
,
\label{gluinomass}
\end{eqnarray}
where $m_{3/2}$ is the gravitino mass, $m_{\tilde{B}}$, $m_{\tilde{W}}$, and $\mgluino$ 
are the physical masses of the Bino, Wino, and gluino, respectively. 
We used the one--loop renormalization group equations
to evolve the running masses up to the scale $\mu = T_R$. 
Here, the numerical coefficients are evaluated at $T_R=10^8$~GeV.
Logarithmic dependences on $T_R$ is omitted for simplicity, 
but they are included in the following numerical calculation.
The reheating temperature $T_R$ 
is defined as $T_R = (\pi^2 g_*(T_R) / 90)^{-1/4}\sqrt{\Gamma_\phi M_P}$,
where $g_*$ is the effective number of massless degrees of freedom. 
In this letter, $m_{\tilde{B}}$ and $m_{\tilde{W}}$ are taken to be free parameters. 
Thus, given the Bino and Wino masses, we obtain upper bounds on the gluino mass
from Eq.~(\ref{gluinomass}) by using the observed dark matter density 
$\Omega_{DM} h^2 \le 0.121$ at $2\sigma$~\cite{Amsler:2008zzb}.

\subsection{BBN constraints}
\label{sec:BBN}

The BBN constraint is represented by the mass, the lifetime and the abundance of the 
stau. For a given mass and abundance of the stau, there is an upper bound on the stau lifetime, 
$\tau_{\tilde \tau} \simeq 48\pi M_{\mathrm P}^2 m_{3/2}^2 m_{\tilde \tau}^{-5}$.
On the other hand, the abundance is determined by the stau mass. 
Thus, once we provide the mass of the stau, we obtain an upper bound on the gravitino mass. 

\subsubsection{stau annihilation via electroweak processes}

The stau abundance is determined not only by the stau mass but also by its interaction 
with the other particles. 
In the SSM, the electroweak processes usually dominate the annihilation rate of the stau. 
The abundance below the freeze-out temperature is then estimated 
as~\cite{Asaka:2000zh,Fujii:2003nr}~\footnote{
Eq.~(\ref{eq:stau-electroweak}) has an ${\cal O}(10\%)$ uncertainty which 
depends on details of the SSM parameters~\cite{Asaka:2000zh,Fujii:2003nr}.
We also neglected the correction from the Sommerfeld enhancement, which is also 
of the order 10\%~\cite{Berger:2008ti}.}
\begin{eqnarray}
 Y_{\stau} \simeq 7 \times 10^{-14} 
 \times
 \left( \frac{m_{\stau}}{100 \rm{GeV}} \right).
 \label{eq:stau-electroweak}
\end{eqnarray}
Here and hereafter, we define $Y_{\stau} = Y_{\stau^+} + Y_{\stau^-} = (n_{\stau}^+ + n_{\stau}^-)/s$,
where $n_{\stau}^{\pm}$ and $s$ are the number density of the stau and the entropy density,
respectively. 
Then, the BBN constraint leads to~\cite{KKMY},
\begin{eqnarray}
 m_{3/2} \lsim 
 \left\{ \begin{array}{ll}
 0.4 \gev - 10 \gev & (100{\rm GeV}<m_{\stau}<450{\rm GeV}) \\
 10 \gev - 20 \gev & (450{\rm GeV}<m_{\stau}<1000{\rm GeV})
\end{array} \right.,
\end{eqnarray}
where the $^6$Li constraint from the stau catalysis~\cite{cbbn1} gives the bound 
for $m_{\stau}<450{\rm GeV}$, while the deuterium constraint from the hadrodissociation 
due to the stau decay dominates for $m_{\stau} > 450{\rm GeV}$~\cite{KKMY}. 

Fig~\ref{fig:stau-gluino}(a) shows the upper bound on the gluino mass for several reheating 
temperatures. Here, since we consider the stau NLSP scenarios, the region of $m_{\tilde g} 
< m_{\stau}$ is excluded. Furthermore, the Bino and Wino masses are set to be close to the 
stau mass, $m_{\tilde B} = m_{\tilde W} = 1.1 \times m_{\tilde \tau}$, which 
provides a conservative bound, since this minimizes the Bino/Wino contributions to the 
gravitino abundance [see Eq.~(\ref{gluinomass})]. Note that if $m_{{\tilde B} ({\tilde W})}$ 
is too degenerate with $m_{\tilde \tau}$, the stau abundance is enhanced by the ${\tilde B}$ 
(${\tilde W})$ decay, resulting in more stringent bounds on $T_R$ (or $m_{\tilde g}$)~\cite{Asaka:2000zh}. 
Although the gluino mass can be less than $m_{\tilde B}$ and $m_{\tilde W}$ above $m_{\tilde \tau}$, 
the region is very narrow, and thus it is discarded in Fig~\ref{fig:stau-gluino}, for simplicity. 

We found that the gluino mass can be as large as about 2.4~TeV (2~TeV) for $m_{\tilde \tau} \leq 
1$~TeV (500~GeV) when the reheating temperature is $T_R \simeq 10^8$GeV.
For a higher reheating temperature, $T_R\gsim  3(5) \times 10^8$GeV, there are upper 
bounds on both the gluino mass and the stau mass, $m_{\tilde g}\lsim 1$~TeV (0.7~TeV) 
and $m_{\tilde \tau}\lsim 750$~GeV (500~GeV). We also notice that there is no allowed 
region for $T_R\ge 10^9$~GeV.

In the figure, the stau mass ranges up to 1~TeV. Although a very high reheating temperature, 
e.g. $T_R\ge 10^9$~GeV, is not allowed in this mass range, there might exist such parameter 
regions if we consider a heavier stau. However, extrapolating the BBN bound in \cite{KKMY}, 
and taking account of the constraint on the gravitino abundance  from the stau decay,
we do not find any parameter space which satisfies $T_R\gsim  3 \times 10^8$GeV for 
$m_{\tilde \tau} > 1$~TeV as long as the electroweak interactions dominate the stau annihilation.

\subsubsection{stau annihilation via enhanced Higgs channel}

The stau annihilation via the Higgs production channel may be enhanced by a large 
Higgsino mass parameter $\mu$ and $\tan\beta$, the ratio of the up- and down-type Higgs 
vacuum expectation values~\cite{enhance1,enhance2}. Actually, the trilinear coupling of the 
lighter stau with the lightest Higgs is
\begin{eqnarray}
 {\cal L} = -{\cal A}_{\stau \stau h} \stau_1^*\stau_1 h,
 \label{eq:stau-trilinear}
\end{eqnarray}
where the coefficient is
\begin{eqnarray}
 {\cal A}_{\stau \stau h} &=& 
 - \frac{g m_\tau}{2 M_W} (\mu\tan\beta + A_\tau) \sin 2\theta_\tau
 + \frac{g m_\tau^2}{M_W}
 \nonumber \\ &&
 - g_Z M_Z \left[
   \left(-\frac{1}{2}+\sin^2\theta_W\right)\cos^2\theta_\tau
   - \sin^2\theta_W \sin^2\theta_\tau
 \right].
\end{eqnarray}
Here, $g$, $g_Z$,  $M_W$, $M_Z$, and $\theta_W$ are the Standard Model parameters. 
$A_{\tau}$ is the stau trilinear coupling with the down-type Higgs boson
and $\theta_\tau$ is the stau mixing angle. We assumed the decoupling 
limit of the heavy Higgs. When $\mu\tan\beta$ is large, the first term in the right-hand side 
becomes enhanced.

The stau relic abundance is obtained by solving the Boltzmann equation.
It is approximately given by
\begin{eqnarray}
 Y_{\stau} \simeq 
 1.0\times10^{-15}
 \left(\frac{10^{-5}{\rm GeV^{-2}}}{\langle \sigma v\rangle}\right)
 \left(\frac{200{\rm GeV}}{m_{\stau}}\right),
 \label{eq:stau-abundance}
\end{eqnarray}
where $\langle \sigma v\rangle$ is the thermally averaged annihilation cross section.
When the electroweak process is dominant, we obtain the result in Eq.~(\ref{eq:stau-electroweak}). 
On the other hand, as $\mu\tan\beta$ increases, the Higgs- and the top-pair production 
channel is enhanced. Then, the cross section becomes 
\begin{eqnarray}
 \langle \sigma v\rangle \simeq 
 \frac{{\cal A}_{\stau \stau h}^4}{64\pi m^6_{\stau}} f_h +
 \frac{3Y_t^2{\cal A}_{\stau \stau h}^2}{128\pi m_{\stau}^4} f_t,
 \label{eq:cross-section}
\end{eqnarray}
where $Y_t \simeq 1$ is the top Yukawa 
coupling, and  $f_h$ and $f_t$, are defined as
\begin{eqnarray}
f_h = \frac{\sqrt{1-r_h}}{(1-r_h/2)^2} \theta(1 - r_h),~~~
f_t = \frac{(1-r_t/2)\sqrt{1-r_t}}{(1-r_h/4)^2} \theta(1 - r_t),
\end{eqnarray}
with $r_h = m_h^2/m_{\tilde \tau}^2$ and $r_t = m_t^2/m_{\tilde \tau}^2$.
In the following, we use $m_t = 173$~GeV and $m_h = 120$~GeV.

However, such a large trilinear coupling ${\cal A}_{\stau \stau h}$ may suffer from 
a disastrous charge/color breaking (CCB) of the Standard Model \cite{enhance1}. 
Actually, the trilinear interaction ${\cal A}_{\stau \stau h}$ can provide another vacuum 
apart from our vacuum in the stau-Higgs field space. As ${\cal A}_{\stau \stau h}$ 
is enhanced, the CCB vacuum becomes deeper than our vacuum, the potential 
barrier between the two vacua is lowered, and hence our vacuum can decay 
into the CCB vacuum. 

In order to avoid the CCB, the lifetime of our vacuum is required to be longer 
than the age of the universe. The vacuum decay rate is evaluated by the bounce 
method~\cite{vacuum1} at the zero temperature, $T = 0$ \cite{enhance1}. 
In addition, the vacuum can transit through thermal effects in the early universe. 
We analyzed the thermal transition by the method explored in Ref.~\cite{vacuum2}. 
The Higgs potential is evalulated at the one-loop level both for $T=0$ and $T>0$.
We also take account of the thermal potential, which comes from the top quark and 
the electroweak gauge bosons. As a result, 
we obtain an upper bound on ${\cal A}_{\stau \stau h}$
as a function of $m_{\tilde\tau}$.\footnote{
  The result is independent of the sign of $\mu$, because if we flip it, 
  the stau mixing angle $\theta_\tau$ changes the sign, too. 
}
Then, by using Eqs.~(\ref{eq:stau-abundance}) and (\ref{eq:cross-section}),
the minimum abundance of the stau is obtained, 
as is shown in Fig.\ref{enhanceYt}. In detail, the constraint comes from the vacuum 
decay at $T = 0$ for $m_{\tilde\tau} \gsim 220\,{\rm GeV}$, while the finite temperature 
effect dominates for $m_{\tilde\tau} \lsim 220\,{\rm GeV}$. We approximated a bounce 
configuration by a straight line in the stau-Higgs field space. 
In fact, we checked at several model points that the approximation is accurate; 
the error is ${\cal O}(1)$\% compared to the full two-dimensional result.

\begin{figure}[h]
\begin{center}
\begin{tabular}{cc}
\includegraphics[scale=0.55]{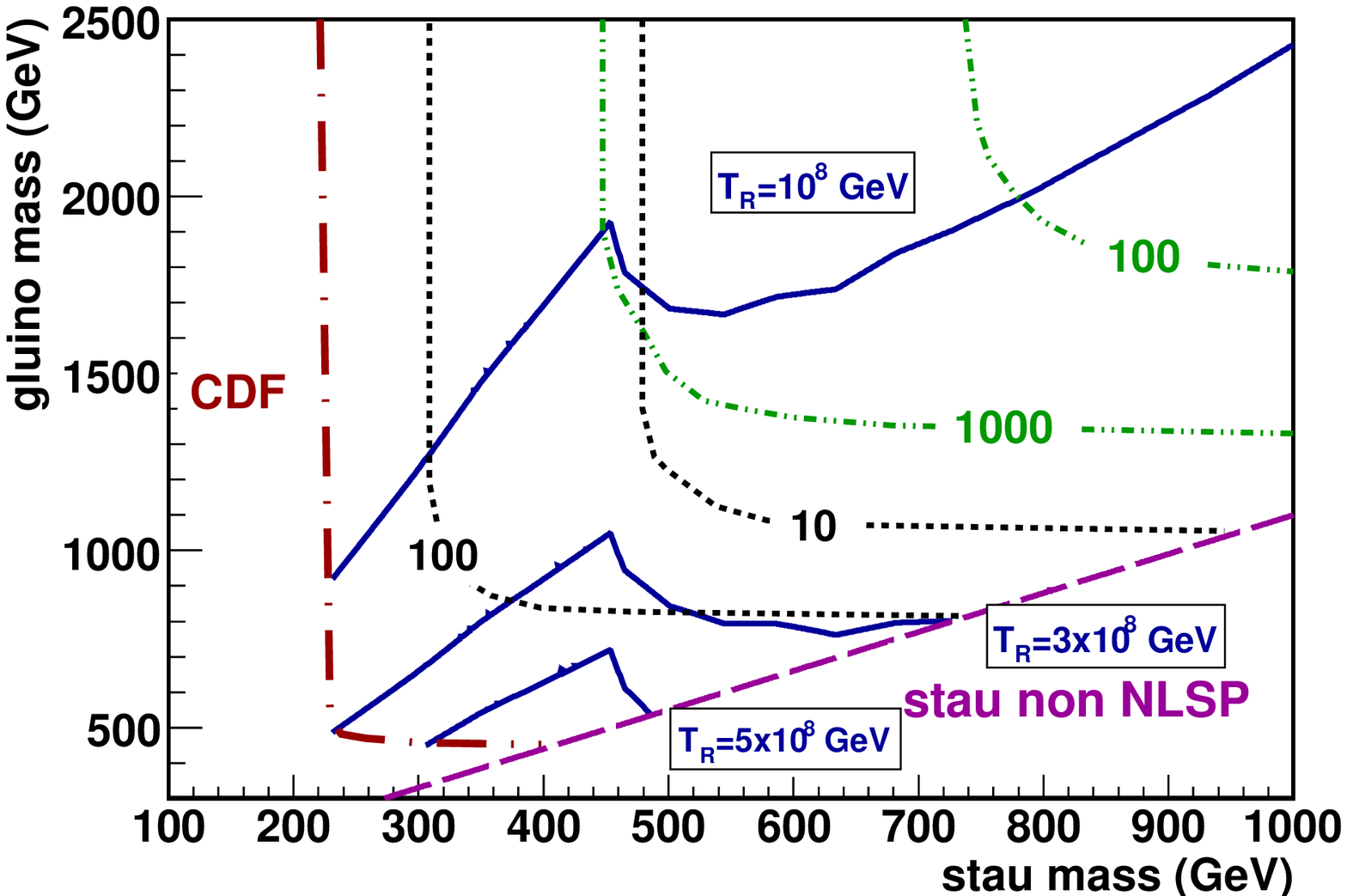} \\ 
(a) \\
\includegraphics[scale=0.55]{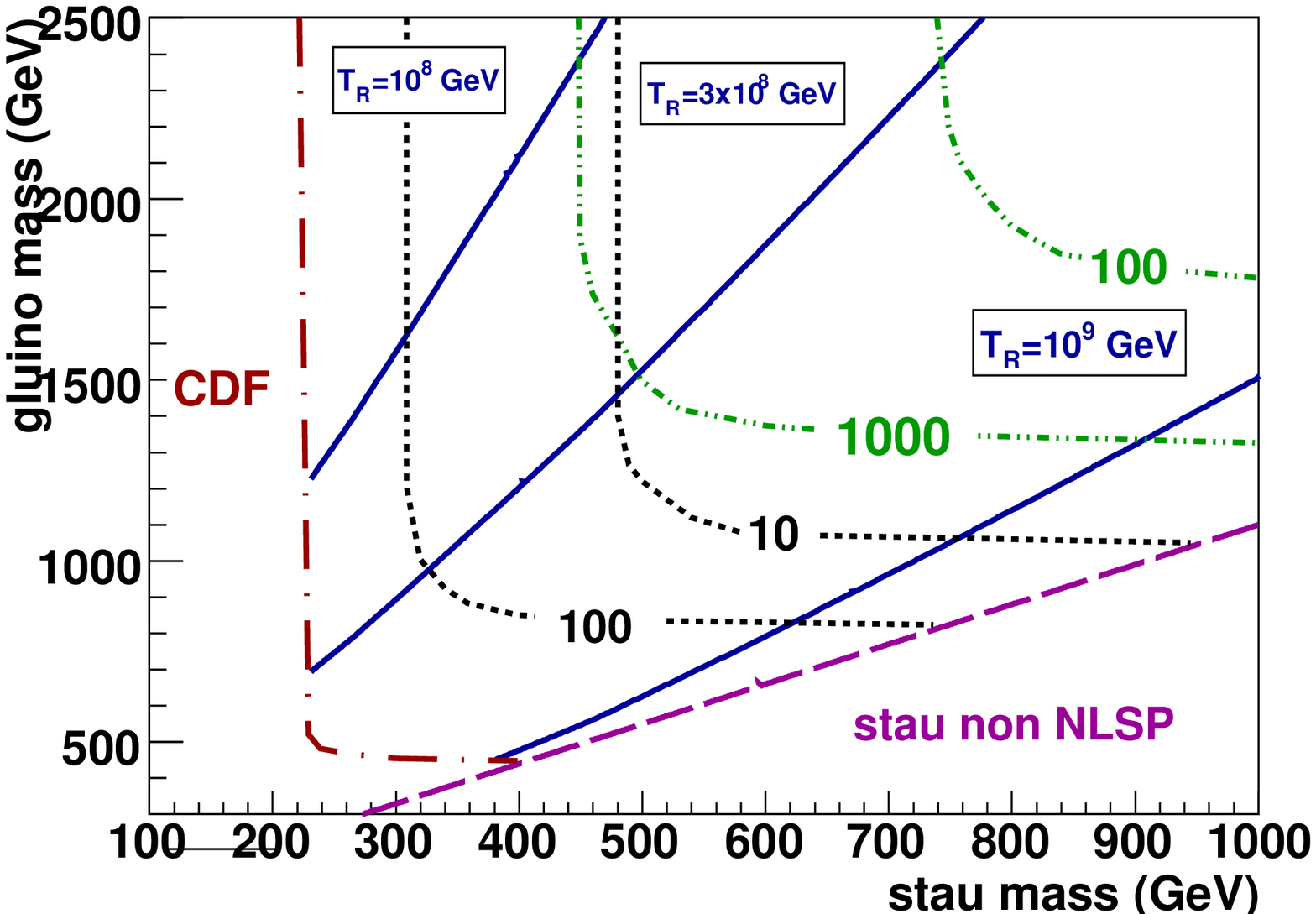} \\ 
(b)
\end{tabular}
\end{center}
\caption{
The solid (blue) lines are upper bounds on the gluino mass for various reheating temperatures 
$T_{R}$ with $m_{\tilde{B}}=m_{\tilde{W}}=1.1~m_{\stau}$. 
The stau annihilation is dominated by (a) the electroweak interactions and (b) the Higgs 
production channel. On the other hand, the dotted (black) and the two-dotted dashed (green) 
lines are contours of the produced stau numbers at the LHC with the integrated luminosity 
1~fb$^{-1}$ at $\sqrt{s}=7$~TeV and 10~fb$^{-1}$ at $\sqrt{s}=14$~TeV, respectively. 
Here, the SUSY events only include the productions of the gluino, chargino and/or neutralino 
(and stau), and no cuts and triggers are imposed. The dotted-dashed (red) line is obtained 
from the CDF bound. Under the dashed (purple) line, the masses satisfy $m_{\tilde g} < 
m_{\tilde{B}} = m_{\tilde{W}} = 1.1~m_{\stau}$.}
\label{fig:stau-gluino}
\end{figure}

\begin{figure}[t]
\begin{center}
\includegraphics[scale=0.6]{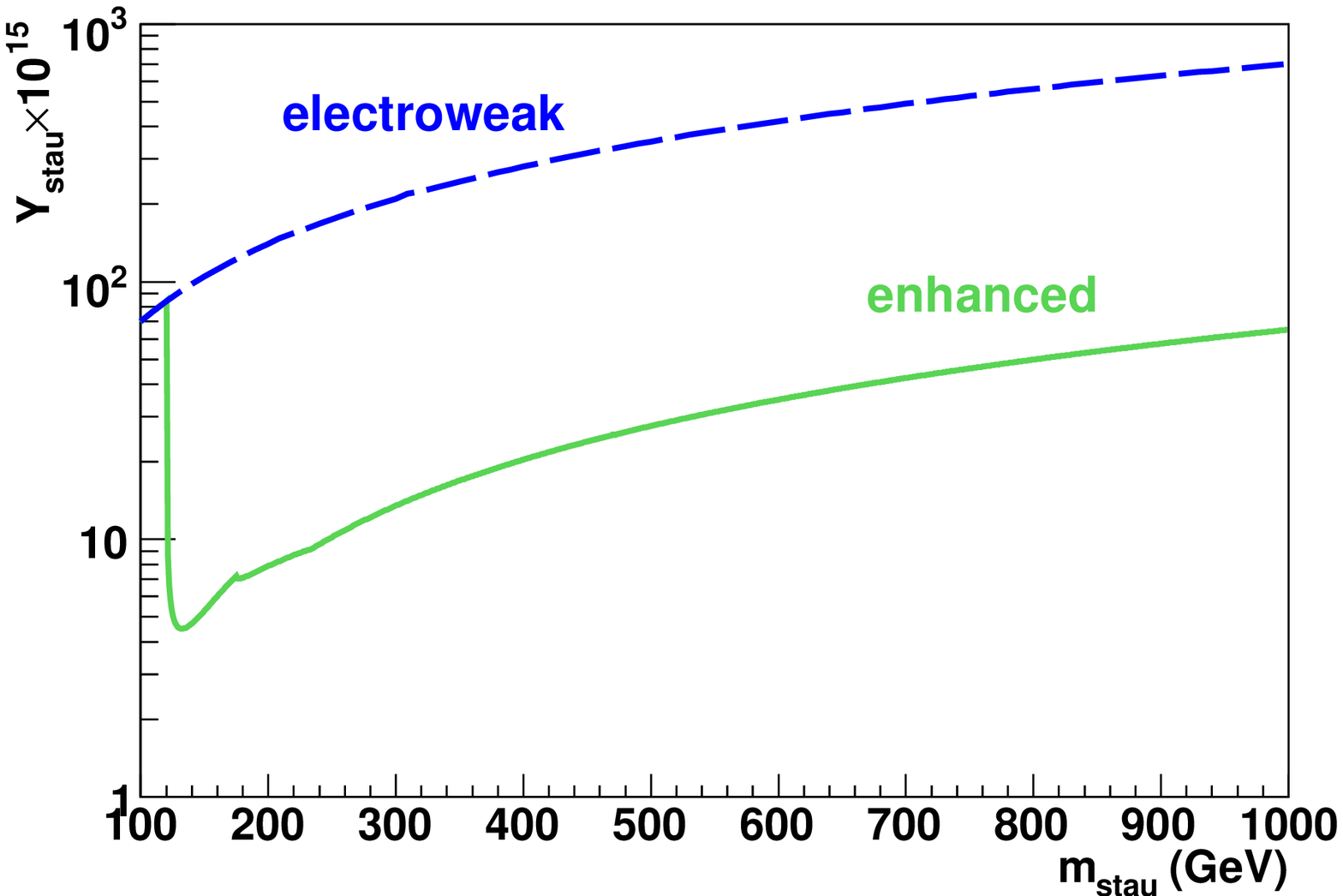}
\caption{The stau abundance $Y_{\stau}$ for varying the stau mass $m_{\stau}$. 
The electroweak interaction gives the stau annihilation on the dashed line, while 
the Higgs (and top) production channel is dominant on the solid line. Here, the stau coupling 
to the Higgs boson, ${\cal A}_{\stau\stau h}$, is taken to be maximal for the latter case. 
The top and Higgs masses are set to be $m_t = 173$~GeV and $m_h = 120$~GeV.}
\label{enhanceYt}
\end{center}
\end{figure}

We found that, although $Y_{\tilde\tau}$ decreases more than one order of magnitude 
compared to the electroweak dominant case, since the stau abundance cannot be less 
than $3 \times 10^{-15}$ for $m_{\tilde \tau} > 100$GeV, the $^6$Li overproduction from 
the stau catalysis still provides the severest constraint. The $^6$Li bound depends 
only on the lifetime and the abundance of the stau, which can be read 
from Fig.~13 of Ref.~\cite{KKMY}. By using the result in Fig.~\ref{enhanceYt}, we obtain
\begin{eqnarray}
 m_{3/2} \lsim \begin{array}{ll}
 0.9 {\rm GeV}-115 \rm{GeV}& (100{\rm GeV}<m_{\stau}<1000 {\rm GeV}).
\end{array} 
\end{eqnarray}
Combined with the constraint from the gravitino abundance in Eq.~(\ref{gluinomass}), 
we obtain upper bounds on the gluino mass  as shown in Fig.~\ref{fig:stau-gluino}(b). 
Note that the gluino bound is similar to that of the electroweak case for $m_{\stau} < 
450{\rm GeV}$, because 
the constraint comes from $^6$Li in both cases, while it is greatly ameliorated for 
$m_{\stau} > 450{\rm GeV}$ due to the absence of deuterium constraint. 
As a result, there is a region where a high reheating temperature, $T_R \gsim 10^9$GeV, 
is realized, which is favored by the thermal leptogenesis~\cite{enhance1,enhance2}.

\section{Collider signatures}
In this section, we study the Tevatron constraints and LHC signatures of the present scenario. 
Since the stau is long-lived, the SUSY events predict charged tracks in the detectors. 
In the following analysis, we focus on the gaugino productions at the $p\bar p$ or $pp$ 
collisions by assuming that the Higgsino and the scalar superparticles other than the stau 
are heavy enough not to contribute to the cross section. Then, the gauginos immediately 
decay into the stau in the end of the decay chains. As long as the squarks have masses of 
${\cal O}(1-10)$TeV, the events are insensitive to the detail of the mass spectrum of the heavy 
superparticles. Furthermore, although the cosmological constraints in the last section 
depend on the size of ${\cal A}_{\stau\stau h}$,  the production cross 
section of the stau is almost independent of it. 

For the numerical analysis, we use PYTHIA 6.4.22 \cite{pythia} to simulate the kinematics 
of the events, and relied on PGS4 \cite{pgs} for the detector simulation. The cross sections 
are estimated at the NLO level by Prospino 2.1 \cite{prospino} for the gluino pair production, 
while the other (non-colored) channels are calculated at the LO by PYTHIA. 

\subsection{Tevatron constraints}
The stable charged massive particles (CHAMPs), i.e. the stau in the present scenario, 
have been already constrained by the Tevatron experiments. Such a long-lived CHAMP looks 
like a heavy muon in the detector. In order for the particle to be distinguished from the muon, 
the experiments have searched for slow CHAMPs with a high transverse momentum. According 
to the CDF result \cite{tevatron}, the events were selected by the following trigger: the highest 
$p_T$ ``muon'' candidate has $p_T > 20$ GeV, satisfying an isolation criteria $\sum 
E_T(0.4)/p_T({\rm muon}) < 0.1$, where $\sum E_T(0.4)$ is the sum of the transverse energy 
within a cone $R=0.4$ around the candidate, excluding the energy which is deposited by the 
candidate itself. The CDF obtained the constraint that the production cross section of the 
non-colored CHAMPs with $|\eta|<0.7$, $p_{{\rm T}} > 40$ GeV and $0.4 < \beta < 0.9$ is 
bounded to be less than 10~${\rm fb}$ at the 95\% C.L.. 

We show the CDF bound in Fig.~\ref{fig:stau-gluino}. When the gluino is relatively light, 
the staus are produced via the gluino channel, while the neutralino/chargino productions 
become the main mode for a heavier gluino. In both channels, almost 
all the passed events are triggered by isolated staus. 
Here, we assumed $m_{\tilde{B}}=m_{\tilde{W}}=1.1~m_{\stau}$ (see Sec.~\ref{sec:BBN}).
As $m_{\tilde{B}}$ and $m_{\tilde{W}}$ increase with $m_{\stau}$ fixed, the CDF bound 
on the stau mass is weakened because the chargino/neutralino production cross section 
decreases, while the cosmological constraint on $m_{\tilde{g}}$ (or $T_R$) becomes 
severer (see Eq.~(\ref{gluinomass})). Although the bound is based on 
the CDF data with the integrated luminosity 1~${\rm fb}^{-1}$, we checked that the 
sensitivity does not change greatly even for $10~{\rm fb}^{-1}$.

\subsection{LHC signatures}
Now let us discuss the LHC signatures.  In the early stage, 
the LHC runs at the beam energy $\sqrt{s}=7~{\rm TeV}$ with an integrated luminosity 
up to about 1~fb$^{-1}$, and it is planned to be upgraded to $\sqrt{s} = 14~{\rm TeV}$. 
In Fig.~\ref{fig:stau-gluino}, we show the contours of the produced stau numbers at the 
LHC for 1~fb$^{-1}$ at $\sqrt{s}=7~{\rm TeV}$ and 10~fb$^{-1}$ at $\sqrt{s}=14~{\rm TeV}$.
Similarly to the Tevatron, the vertical line comes dominantly from the pair production 
of the lightest charginos or that of the lightest chargino and the second lightest neutralino, 
while the horizontal line is obtained by the gluino production. 
As in the Tevatron case, if $m_{\tilde{B}}$ and $m_{\tilde{W}}$ are increased
from  $m_{\tilde{B}}=m_{\tilde{W}}=1.1~m_{\stau}$,
the vertical lines move to the left (i.e., less staus are produced), while
the bound on $m_{\tilde{g}}$ (or $T_R$) becomes severer.
It should be mentioned that 
we imposed no cuts and triggers, which will be studied below. We find that much more 
staus will be produced even in the early stage of the LHC compared to the Tevatron. 

The realistic analysis includes the trigger criteria and the cuts to reduce the backgrounds. 
The particles produced in the decay chain can be energetic enough to trigger the events. 
Moreover, the stau itself may hit the muon trigger. In this letter, we adopt the following 
trigger menu, 
\begin{itemize}
\item at least one isolated electron has $p_{\rm T}>20$~GeV,
\item at least one isolated muon has $p_{\rm T}>40$~GeV,
\item at least one isolated tau has $p_{\rm T}>100$~GeV,
\item at least one isolated stau has $p_{\rm T}>40$~GeV within the bunch,
\item at least two staus have $p_{\rm T}>40$~GeV within the bunch.
\end{itemize}
If any one of these conditions is satisfied, the event is read out. Here, the isolation condition 
relies on PGS4, but the muon and the stau are considered to be isolated if they satisfy 
the following two conditions: 1) the summed $p_{{\rm T}}$ in a $R=0.4$ cone around the 
particle (excluding the particle itself) is less than 5 GeV, and 2) the ratio of $E_{{\rm T}}$ in 
a $3\times3$ calorimeter array around the particle (including the particle's cell) to the 
$p_{{\rm T}}$ of the particle is less than 0.1125. In the last two trigger items, we imposed the 
bunch condition, because the stau trigger works only when the stau reaches the muon trigger 
before the next bunch collides. Thus, the stau velocity is required to be $\beta>0.7$ in the 
barrel region, $|\eta|<1.0$, and $\beta>0.8$ in the endcap region, $1.0<|\eta|<2.8$, for the 
trigger to be applied \cite{Aad:2009wy}. Although we set relatively high thresholds for the 
triggers, the event number is almost insensitive to them. In addition to the above menu, one may 
take the jet triggers into account. We checked that this does not change the number of events 
for $\sqrt{s}=7~{\rm TeV}$, while it can increase by an ${\cal O}(1)$ factor for $\sqrt{s}=14~{\rm TeV}$.

After selecting the events based on the above trigger menu, the stau number is counted with 
the following cut conditions for the stau:
\begin{itemize}
\item $p_{{\rm T}}>20~{\rm GeV}$
\item $0.5<\beta<0.9$
\item $|\eta|<2.5$.
\end{itemize}
Actually, the first two cuts significantly reject the muon background, because a high $p_T$ 
muon has a velocity of $\beta \simeq 1$. The last one is imposed for the stau 
to be detected in the muon detector \cite{Aad:2009wy,reconstruction}. 

The number of events may be further reduced by taking the efficiency into account. In the 
current setup, a large part of the events passes the muon (stau) triggers. In fact, the most 
events are triggered by the stau when the chargino/neutralino is the dominant channel. 
In the gluino production case, roughly a half of the total passed events is accepted by the 
stau trigger, and the others are by the leptons. Since the trigger efficiency of the muon-like 
event is pretty good \cite{Aad:2009wy}, the reduction is expected to be small. 
On the other hand, the reconstruction efficiency of the stau depends on the LHC analysis. 
According to the ATLAS CSC studies \cite{Aad:2009wy}, the efficiency varies from 0.1 to 
0.9 for $\beta \simeq 0.5 - 0.9$. 
The analysis can be improved by the method studied in \cite{reconstruction}, which provides 
the efficiency more than 90\% for $\beta \gsim 0.5$. In the following result we assume the 
efficiency to be 100\% for simplicity. 

\begin{figure}[h]
\begin{center}
\begin{tabular}{cc}
\includegraphics[scale=0.6]{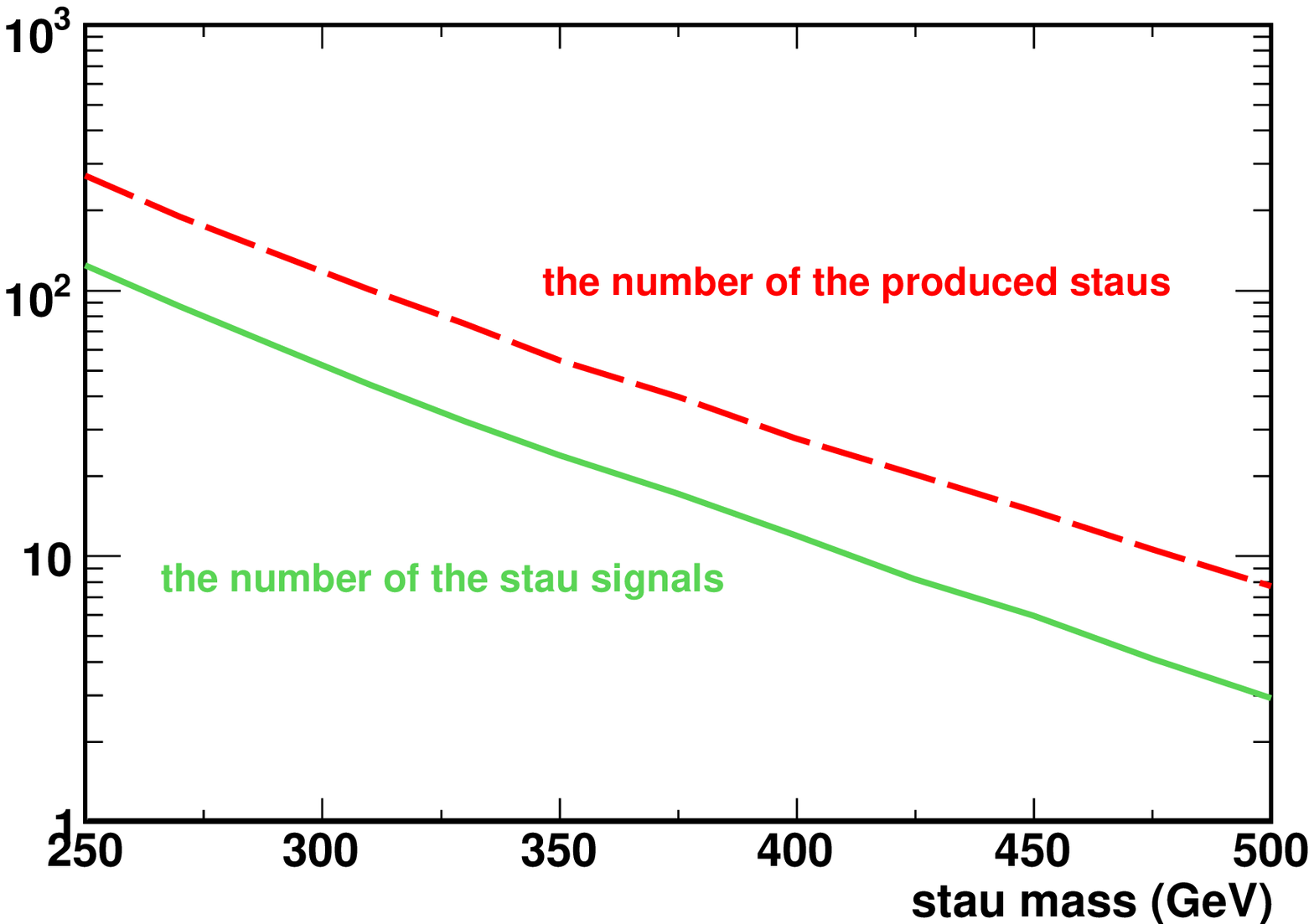} \\ 
(a) \\
\includegraphics[scale=0.6]{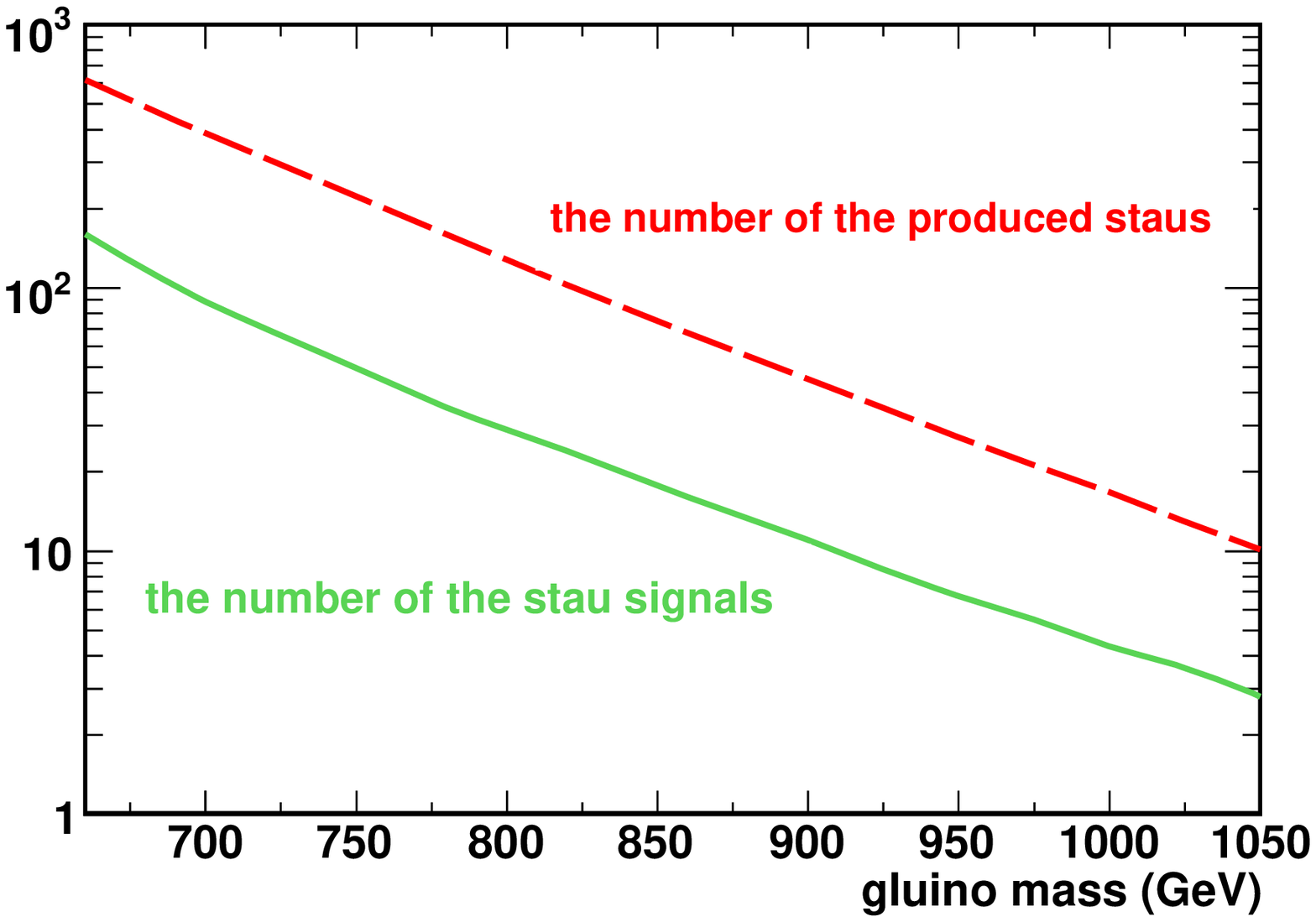} \\ 
(b)
\end{tabular}
\end{center}
\caption{The solid line shows the number of the generated staus after applying 
the triggers and the cuts, which is compared to the stau number without any 
conditions displayed by the dashed line. The gluino mass is set to be 2TeV for (a), 
and the stau mass is 600GeV for (b). The beam energy is chosen to be 
$\sqrt{s}=7$~TeV. }
\label{fig:aftercut}
\end{figure}

In Fig.~\ref{fig:aftercut}, we show the number of the staus for the LHC beam energy $\sqrt{s}
=7$~TeV and an integrated luminosity 1 fb$^{-1}$. The result after taking account of the 
detector simulation is compared with that without any cut conditions. We find that the number 
decreases down to $\sim 50\%$ and $20-30\%$ for Fig.~\ref{fig:aftercut}(a) and (b), respectively. 
More precisely, the efficiency depends on details of the superparticle mass 
spectrum.

The long-lived CHAMP events are almost free from the background, and therefore 
they can be discovered with a small number of the events. 
For instance, we see that the LHC will produce more than 5 (10) staus for $m_{\stau} 
\lsim 460$GeV (410GeV) when the chargino/neutralino production is dominant 
(Fig.~\ref{fig:aftercut}(a)), and the same for $m_{\tilde g} \lsim 990$GeV (910GeV) 
in the case of the gluino production (Fig.~\ref{fig:aftercut}(b)). 
In a smaller mass region, the stau number depends on both the stau and 
gluino masses. Especially, we checked that the whole region with $T_R \gsim 3 \times 
10^8$GeV  in 
Fig.~\ref{fig:stau-gluino}(a) predicts more than 10 staus after applying the cut conditions. 
Furthermore, when the LHC beam energy is upgraded to $\sqrt{s}=14$~TeV and the 
integrated luminosity reaches 10 fb$^{-1}$, the sensitivity becomes greatly improved. 
Actually, we checked that more than 10 staus can be detected even for $m_{\stau} = 1$TeV 
and $m_{\tilde g} = 2.5$TeV. 

In the above analysis, we assumed that the scalar superparticles, particularly the 
squarks, are very heavy. As the squarks become lighter, productions of the squarks 
come into the channels. In fact, when the squark has a mass comparable to the gluino 
mass, the squark channels become dominant, and the SUSY cross section is enhanced 
by one order of magnitude compared to that of the gluino. Thus, it is considered that the 
result in this letter is conservative, and we may observe more SUSY events even in the 
early stage of the LHC. 

\section{Conclusion and Discussions}
In this letter, we studied the cosmology and the collider features of 
high reheating temperature scenarios with a long-lived stau
within the SSM. We first discussed the cosmological constraint from 
the gravitino abundance and the BBN. In most of the parameter space, the stau 
annihilation is dominated by the electroweak interactions, and we obtained the upper 
bounds on the gluino mass
 for a high reheating temperature. When the reheating temperature is 
$T_R \gsim 10^8$GeV, the gluino is limited to be lighter than about 2.4~TeV for 
$m_{\tilde \tau} \leq 1$TeV. A higher reheating temperature provides a stronger 
constraint, $m_{\tilde g}\lsim 1$~TeV (0.7~TeV) and $m_{\tilde \tau}\lsim 750$~GeV 
(500~GeV) for $T_R\gsim 3(5) \times 10^8$GeV. In addition, there is no allowed region 
for $T_R \ge 10^9$~GeV. We then investigated the collider signatures. We found that 
more than 10 staus will be detected for $T_R\gsim 3 \times 10^8$GeV in the early 
LHC for an integrated luminosity $1~{\rm fb}^{-1}$ with $\sqrt{s}=7$~TeV. 
Upgrading the LHC to $\sqrt{s} = 14~{\rm TeV}$, we expect $\gsim 10$ staus in the whole 
mass region of $T_R\gsim 10^8$GeV with $m_{\tilde \tau} \leq 1$TeV for an integrated 
luminosity $10~{\rm fb}^{-1}$.

When the Higgs coupling to the stau is very large, the cosmological constraint becomes 
ameliorated, especially for a heavy stau. Interestingly, this opens a window for a high 
reheating temperature, $T_R \gsim 10^9$GeV for $m_{\stau} \gsim 400$GeV, which saves 
the thermal leptogenesis~\cite{enhance1,enhance2}. We found that this region predicts 
more than 5 staus in the early LHC if the stau mass is $m_{\stau} \lsim 700$GeV, and the 
stau number becomes much more than 10 for $m_{\stau} \lsim 1$TeV till the end 
of the LHC. Thus, the thermal leptogenesis in the present framework can be tested
by the LHC in a wide parameter region of the SSM. 

If the long-lived staus are indeed produced at the LHC, the masses of the stau and other
superparticles can be measured with a good accuracy~\cite{Ito:2009xy}.
It then becomes possible to obtain a more precise (severer) bound on the reheating 
temperature. Furthermore, the stau lifetime (and hence the gravitino mass) may also be 
measured~\cite{Asai:2009ka}. The stau lifetime is directly subject to the BBN constraint, 
while the gravitino mass gives a further information on the gravitino abundance.
In particular, if there is no allowed region in the parameter space of the stau lifetime and the 
stau abundance, it indicates a very low reheating temperature~\cite{Takayama:2007du} 
or an entropy production between stau decoupling and the BBN~\cite{Buchmuller:2006tt}.

Although we studied the electroweak and the Higgs production channels of the stau 
annihilation, it is difficult to distinguish them merely by counting the stau number at the LHC. 
This is because the event kinematics is almost insensitive to the stau--Higgs interaction. 
Noting that the lighter stau is likely to mix with both the left- and right-handed staus maximally 
in order to realize a large ${\cal A}_{\stau\stau h}$, we may measure the mixing angle by determining 
the chirality structure of the stau, for example, by measuring a polarization of the tau which 
is associated with the stau production \cite{Kitano:2010tt}. 

In this letter, we discussed the stau annihilation via the electroweak interactions and 
the light Higgs channel. Ref.~\cite{enhance2} also studied a possible annihilation channel 
of the stau through the heavy Higgs exchange. If the masses are tuned well such that the 
stau annihilates through the heavy Higgs pole, the stau abundance may decrease considerably. 
This scenario may be interesting from the viewpoint of the LHC signatures, since the heavy 
Higgs has a strong interaction with the stau and its mass is correlated with the stau mass.

When the stau coupling to the Higgs is strong, the SUSY contribution to the muon anomalous 
magnetic moment may become large. This is because large $\mu$ and $\tan\beta$ enhance 
the mixing of the left- and right-handed sleptons, namely the smuon--neutralino contribution to 
the muon $g-2$ \cite{Moroi:1995yh}. Since the experimental result \cite{Bennett:2006fi} shows 
a 3--4$\sigma$ discrepancy from the SM prediction~\cite{Hagiwara:2006jt}, 
a large ${\cal A}_{\stau\stau h}$ might be an interesting possibility. 

\section*{Acknowledgment}
We would like to thank J.~Hisano, T.~Ito, S.~Matsumoto, T.~Moroi, S.~Shirai for useful comments.
The work of K.H. was supported by JSPS Grant-in-Aid for Young Scientists (B) (21740164) and
Grant-in-Aid for Scientific Research (A) (22244021).
This work was supported by World Premier International Center Initiative (WPI Program),
MEXT, Japan.


\begin{thebibliography}{99}

\bibitem{TR5MeV}
 M.~Kawasaki, K.~Kohri and N.~Sugiyama,
 Phys.\ Rev.\ Lett.\  {\bf 82} (1999) 4168
 [arXiv:astro-ph/9811437];
 Phys.\ Rev.\  D {\bf 62} (2000) 023506
 [arXiv:astro-ph/0002127];
 S.~Hannestad,
 Phys.\ Rev.\  D {\bf 70} (2004) 043506
 [arXiv:astro-ph/0403291];
 K.~Ichikawa, M.~Kawasaki and F.~Takahashi,
 Phys.\ Rev.\  D {\bf 72} (2005) 043522
 [arXiv:astro-ph/0505395].

\bibitem{Nakayama:2008wy}
 K.~Nakayama, S.~Saito, Y.~Suwa and J.~Yokoyama,
 JCAP {\bf 0806} (2008) 020
 [arXiv:0804.1827 [astro-ph]].

\bibitem{Takayama:2007du}
  F.~Takayama,
  Phys.\ Rev.\  D {\bf 77} (2008) 116003
  [arXiv:0704.2785 [hep-ph]].


\bibitem{Asaka:1999jb}
 T.~Asaka, K.~Hamaguchi, M.~Kawasaki and T.~Yanagida,
 Phys.\ Rev.\  D {\bf 61} (2000) 083512
 [arXiv:hep-ph/9907559];

\bibitem{Endo:2007sz}
 M.~Endo, F.~Takahashi and T.~T.~Yanagida,
 Phys.\ Rev.\  D {\bf 76} (2007) 083509
 [arXiv:0706.0986 [hep-ph]].


\bibitem{FY}
 M.~Fukugita and T.~Yanagida,
 Phys.\ Lett.\  B {\bf 174} (1986) 45.



\bibitem{Lepto_review}
 W.~Buchmuller, R.~D.~Peccei and T.~Yanagida,
 Ann.\ Rev.\ Nucl.\ Part.\ Sci.\  {\bf 55} (2005) 311
 [arXiv:hep-ph/0502169];
 S.~Davidson, E.~Nardi and Y.~Nir,
 Phys.\ Rept.\  {\bf 466} (2008) 105
 [arXiv:0802.2962 [hep-ph]].

\bibitem{KKMY}
 M.~Kawasaki, K.~Kohri, T.~Moroi and A.~Yotsuyanagi,
 Phys.\ Rev.\  D {\bf 78}, 065011 (2008)
 [arXiv:0804.3745 [hep-ph]].



\bibitem{Choi:2007rh}
  K.~Y.~Choi, L.~Roszkowski and R.~Ruiz de Austri,
  JHEP {\bf 0804} (2008) 016
  [arXiv:0710.3349 [hep-ph]].

\bibitem{Steffen:2008bt}
  F.~D.~Steffen,
  Phys.\ Lett.\  B {\bf 669} (2008) 74
  [arXiv:0806.3266 [hep-ph]].


\bibitem{Feng:2004mt}
  J.~L.~Feng, S.~Su and F.~Takayama,
  Phys.\ Rev.\  D {\bf 70} (2004) 075019
  [arXiv:hep-ph/0404231];
 J.~L.~Feng, S.~Su and F.~Takayama,
  Phys.\ Rev.\ Lett.\  {\bf 96} (2006) 151802
  [arXiv:hep-ph/0503117].



\bibitem{Fujii:2003nr}
M.~Fujii, M.~Ibe and T.~Yanagida,
Phys.\ Lett.\  B {\bf 579}, 6 (2004)
[arXiv:hep-ph/0310142].

\bibitem{Pradler:2006qh}
 J.~Pradler and F.~D.~Steffen,
 Phys.\ Rev.\  D {\bf 75} (2007) 023509
 [arXiv:hep-ph/0608344].

\bibitem{Ros}
See also,
  L.~Roszkowski, R.~Ruiz de Austri and K.~Y.~Choi,
  JHEP {\bf 0508} (2005) 080
  [arXiv:hep-ph/0408227];






\bibitem{Moroi:1993mb}
 T.~Moroi, H.~Murayama and M.~Yamaguchi,
 Phys.\ Lett.\  B {\bf 303}, 289 (1993);

\bibitem{Bolz:2000fu}
 M.~Bolz, A.~Brandenburg and W.~Buchmuller,
 Nucl.\ Phys.\  B {\bf 606}, 518 (2001)
 [Erratum-ibid.\  B {\bf 790}, 336 (2008)]
 [arXiv:hep-ph/0012052];

\bibitem{Pradler:2006hh}
J.~Pradler and F.~D.~Steffen,
Phys.\ Lett.\  B {\bf 648}, 224 (2007)
[arXiv:hep-ph/0612291].


\bibitem{Rychkov:2007uq}
  V.~S.~Rychkov and A.~Strumia,
  Phys.\ Rev.\  D {\bf 75} (2007) 075011
  [arXiv:hep-ph/0701104].


\bibitem{Amsler:2008zzb}
 C.~Amsler {\it et al.}  [Particle Data Group],
 Phys.\ Lett.\  B {\bf 667} (2008) 1.



\bibitem{Asaka:2000zh}
 T.~Asaka, K.~Hamaguchi and K.~Suzuki,
 Phys.\ Lett.\  B {\bf 490}, 136 (2000)
 [arXiv:hep-ph/0005136].


\bibitem{Berger:2008ti}
 C.~F.~Berger, L.~Covi, S.~Kraml and F.~Palorini,
 JCAP {\bf 0810} (2008) 005
 [arXiv:0807.0211 [hep-ph]].



\bibitem{cbbn1}
M.~Pospelov,
Phys.\ Rev.\ Lett.\  {\bf 98}, 231301 (2007)
[arXiv:hep-ph/0605215].

\bibitem{enhance1}
M.~Ratz, K.~Schmidt-Hoberg and M.~W.~Winkler,
JCAP {\bf 0810}, 026 (2008)
[arXiv:0808.0829 [hep-ph]].

\bibitem{enhance2}
J.~Pradler and F.~D.~Steffen,
Nucl.\ Phys.\  B {\bf 809}, 318 (2009)
[arXiv:0808.2462 [hep-ph]].


\bibitem{vacuum1}
  S.~R.~Coleman,
  Phys.\ Rev.\  D {\bf 15}, 2929 (1977)
  [Erratum-ibid.\  D {\bf 16}, 1248 (1977)];
  \\
  C.~G.~.~Callan and S.~R.~Coleman,
  Phys.\ Rev.\  D {\bf 16}, 1762 (1977).


\bibitem{vacuum2}
  A.~D.~Linde,
  Nucl.\ Phys.\  B {\bf 216}, 421 (1983)
  [Erratum-ibid.\  B {\bf 223}, 544 (1983)].

\bibitem{pythia}
T.~Sjostrand, S.~Mrenna and P.~Z.~Skands,
JHEP {\bf 0605}, 026 (2006)
[arXiv:hep-ph/0603175].
\bibitem{pgs}
The information on Pretty Good Simulation of high energy collisions (PGS4) can be seen in
http://www.physics.ucdavis.edu/\verb|%|7Econway/research/research.html.
\bibitem{prospino}
 W.~Beenakker, R.~Hopker and M.~Spira,
 arXiv:hep-ph/9611232.


\bibitem{tevatron}
 T.~Aaltonen {\it et al.}  [CDF Collaboration],
 Phys.\ Rev.\ Lett.\  {\bf 103}, 021802 (2009)
 [arXiv:0902.1266 [hep-ex]].


\bibitem{Aad:2009wy}
  G.~Aad {\it et al.}  [The ATLAS Collaboration],
  arXiv:0901.0512 [hep-ex].
\bibitem{reconstruction}
S.~Tarem, S.~Bressler, H.~Nomoto and A.~Di Mattia,
Eur.\ Phys.\ J.\  C {\bf 62}, 281 (2009).

\bibitem{Ito:2009xy}
See, for instance, 
  T.~Ito, R.~Kitano and T.~Moroi,
  JHEP {\bf 1004} (2010) 017
  [arXiv:0910.5853 [hep-ph]],
  and references therein.


\bibitem{Asai:2009ka}
See, for instance, 
  S.~Asai, K.~Hamaguchi and S.~Shirai,
  Phys.\ Rev.\ Lett.\  {\bf 103} (2009) 141803
  [arXiv:0902.3754 [hep-ph]],
  and references therein.

\bibitem{Buchmuller:2006tt}
  W.~Buchmuller, K.~Hamaguchi, M.~Ibe and T.~T.~Yanagida,
  Phys.\ Lett.\  B {\bf 643} (2006) 124
  [arXiv:hep-ph/0605164].

\bibitem{Kitano:2010tt}
  R.~Kitano and M.~Nakamura,
  arXiv:1006.2904 [hep-ph].

\bibitem{Moroi:1995yh}
  T.~Moroi,
  Phys.\ Rev.\  D {\bf 53}, 6565 (1996)
  [Erratum-ibid.\  D {\bf 56}, 4424 (1997)]
  [arXiv:hep-ph/9512396].

\bibitem{Bennett:2006fi}
  G.~W.~Bennett {\it et al.}  [Muon G-2 Collaboration],
  Phys.\ Rev.\  D {\bf 73}, 072003 (2006)
  [arXiv:hep-ex/0602035].

\bibitem{Hagiwara:2006jt}
  K.~Hagiwara, A.~D.~Martin, D.~Nomura and T.~Teubner,
  Phys.\ Lett.\  B {\bf 649}, 173 (2007)
  [arXiv:hep-ph/0611102];
  T.~Teubner, K.~Hagiwara, R.~Liao, A.~D.~Martin and D.~Nomura,
  arXiv:1001.5401 [hep-ph];
  M.~Davier, A.~Hoecker, B.~Malaescu, C.~Z.~Yuan and Z.~Zhang,
  Eur.\ Phys.\ J.\  C {\bf 66}, 1 (2010)
  [arXiv:0908.4300 [hep-ph]];

\end{thebibliography}
\end{document}